\documentclass[sigconf]{acmart}

\copyrightyear{2026}
\acmYear{2026}
\setcopyright{cc}
\setcctype{by-nc-nd}
\acmConference[IGSC 2026]{International Green and Sustainable Computing Conference}{June 22--24, 2026}{Canandaigua, NY, USA}
\acmBooktitle{International Green and Sustainable Computing Conference (IGSC 2026), June 22--24, 2026, Canandaigua, NY, USA}
\acmDOI{10.1145/3797248.3816189}
\acmISBN{979-8-4007-2520-3/2026/06}

\begin{document}

\title{To Overlay or to Customize? \\Revisiting Architectural Choices in Heterogeneous Systems}

\settopmatter{authorsperrow=4}
\author{Xingzhen Chen*}
\affiliation{%
  \institution{Brown University} 
  \city{Providence}
  \country{USA}
  }
\email{xingzhen_chen@brown.edu}

\author{Shixin Ji*}
\affiliation{%
  \institution{Brown University} 
  \city{Providence}
  \country{USA}
  }
\email{shixin_ji@brown.edu}

\author{Zheng Dong}
\affiliation{%
  \institution{Wayne State University}
  \city{Detroit}
  \country{USA}
  }
\email{dong@wayne.edu}

\author{Peipei Zhou}
\affiliation{%
  \institution{Brown University} 
  \city{Providence}
  \country{USA}
  }
\email{peipei_zhou@brown.edu}

\renewcommand{\shortauthors}{
Xingzhen Chen,
Shixin Ji,
Zheng Dong,
and Peipei Zhou
}


\begin{abstract}

Autonomous Driving Systems (ADS) increasingly rely on diverse deep neural networks to support perception, prediction, planning, and control under strict real-time constraints. FPGA-based heterogeneous computing provides an attractive platform for DNN workloads, but it raises a fundamental deployment question: should the system rely on a flexible overlay architecture, or repeatedly load customized bitstreams optimized for dedicated models, which should be treated as a first-class systems problem rather than a purely architectural one? Overlay-based execution offers fast model switching and better adaptability relying on lightweight instruction or parameter updates, while customized architectures can provide higher model-wise efficiency at the cost of reconfiguration latency and reduced flexibility. However, the boundary between these two design choices remains unclear in realistic ADS scenarios.

In this work, we present a systematic study of this trade-off from a deployment-centric perspective, focusing on an autonomous driving scenario. Instead of treating overlay and customized acceleration as isolated design points, we analyze when each approach is preferable under practical conditions, including workload variation, architectural design, reconfiguration latency, and switching frequency. Our analysis shows that overlay-based architecture is more suitable for highly frequent model switching under the state-of-the-art architecture. However, as bitstream reload overhead continues to reduce, customized architectures may become increasingly attractive, especially for workloads with efficiency requirements. Conversely, if overlay architectures become more capable and flexible, they may further expand their advantage over customized architectures. These observations provide design insights for future architectural design, and the optimal deployment strategy will be flipped according to the technique development.


\end{abstract}


\maketitle

\begingroup
\renewcommand\thefootnote{}
\footnotetext{*Both authors contributed equally to this research.}
\endgroup

\vspace{-5pt}
\section{Introduction}

Autonomous Driving Systems (ADS) increasingly rely on a diverse set of deep neural networks (DNNs) to support perception, prediction, planning, and control under strict real-time constraints. 
Modern ADS workloads usually consist of multiple DNN tasks with different model structures and sizes, process frequencies, and latency requirements. For example, an autonomous vehicle needs to periodically execute image segmentation, object classification, point-cloud processing, and other DNN models with varying process rates, and each instance must finish before its next release to avoid sensor data backlog and unbounded response latency. Therefore, ADS deployment is not only a DNN acceleration problem, but also a real-time systems problem. 

\begin{figure}
    \centering
    \includegraphics[width=0.8\linewidth]{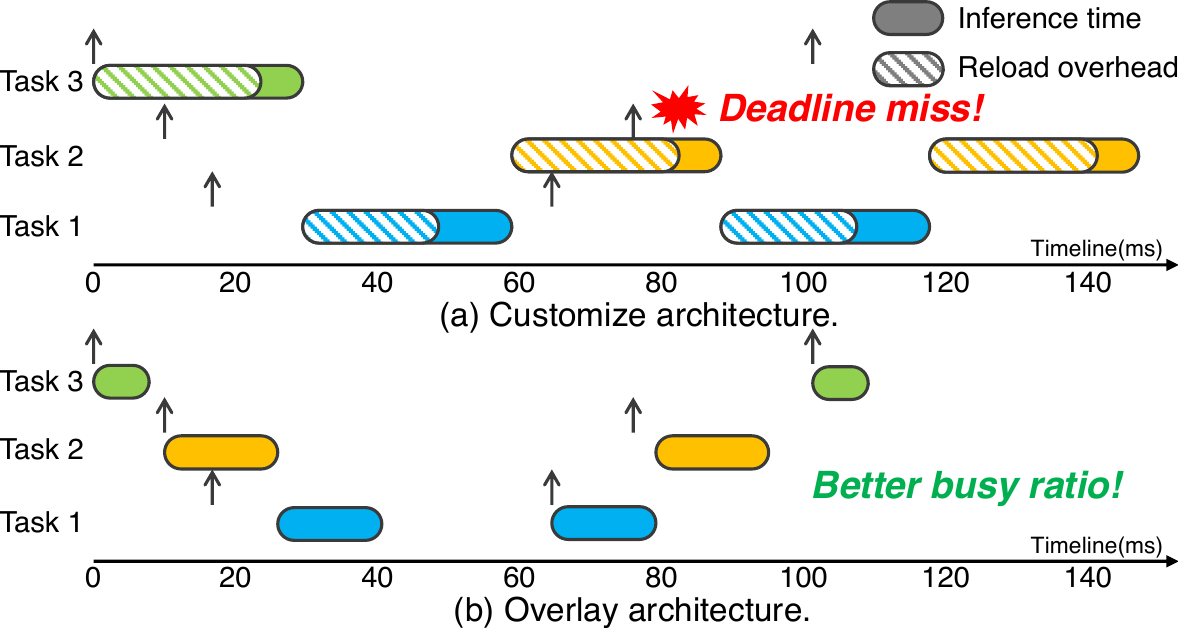}
    \vspace{-10pt}

    \caption{The timeline comparison on customized bitstream and overlay architecture strategies for Setup A.}
    \vspace{-10pt}
    \label{fig: sota_timeline}
\end{figure}

FPGA-based heterogeneous accelerator provides an attractive opportunity for ADS workloads since it has high energy efficiency, architectural flexibility, and the ability to customize computation and data movement for different models.
However, flexibility also raises a fundamental deployment question: should an ADS platform rely on a flexible overlay architecture that one generic architecture handles all DNN workloads, or repeatedly load customized bitstreams that are well-optimized for each task model? 
Hardware customization can exploit model-wise optimization opportunities, such as specific dataflow, kernel fusion, and on-chip data forwarding, thereby improving execution efficiency for dedicated models. 
However, switching between customized accelerators requires bitstream reloading, which introduces non-negligible latency and significantly affects real-time schedulability.
As shown in Figure \ref{fig: sota_timeline}(a), the striped bar represents the bitstream reloading overhead, which is non-negligible compared with execution time, especially in ADS scenarios that have a high-frequency model switching.
In contrast, overlay-based accelerators use a generic bitstream and support different models through lightweight instructions or word-level reconfiguration. 
This enables fast model switching and avoids expensive bit-level reconfiguration, but the generic architecture may lose efficiency compared with model-specific designs. As shown in Figure \ref{fig: sota_timeline}(b), though generality induces throughput drops, the overlay can save overhead on bitstream reloading, thereby performing better in ADS scenarios.
This tradeoff is especially important in ADS scenarios. 
Unlike offline or batch-processing applications, ADS workloads are periodic, latency-sensitive, and frequent task switching. For example, Figure \ref{fig: sota_timeline}(a) shows that the customization strategy would lead to a deadline miss on Task 2, which might result in catastrophic consequences, while Figure \ref{fig: sota_timeline}(b) shows that each task can be completed before the release time of the next instance and with a better accelerator busy ratio, though generic overlay architecture induce throughput drops.
Therefore, should we choose overlay or customized architecture in ADS scenarios, this question should be treated as a first-class systems problem rather than a purely architectural one.

In this work, we present a systematic study of the overlay-customization tradeoff from a deployment-centric perspective, using ADS as a representative real-time application scenario. 
We construct a periodic ADS task model inspired by practical perception workloads, where multiple DNN tasks are released at different sensor frequencies and scheduled on a single computing platform using the Earliest Deadline First (EDF) algorithm. 
We compare two execution strategies: overlay-based strategy that executes all models on a single generic overlay architecture without bitstream reload, and a customized strategy that executes each model using a dedicated bitstream and pays reconfiguration overhead when switching architectures. 
Instead of comparing model-wise end-to-end throughput, we evaluate the end-to-end accelerator busy ratio and schedulability under different workload setups. We also summarize the key takeaways as follows:
\begin{itemize}
\item The optimal architecture for one task is not necessarily the best architecture for the whole system, since bitstream switching makes the system-level performance worse.
\item Reconfiguration overhead is a significant factor in ADS deployment. In ADS workloads with millisecond-level deadlines, even a few milliseconds of reload latency can significantly affect schedulability.
\item Customized architectures might become attractive when bitstream reload overhead is sufficiently reduced.
\item Small models are more sensitive to bitstream reloading overhead due to the small execution time.
\item Advanced overlays can reverse the trend again. If overlay architectures support better data movement and on-chip forwarding for small workloads, they can close the efficiency gap with customized accelerators while keeping their switching advantage.

\end{itemize}

\vspace{-7pt}
\section{Related Works}
\label{sec: related works}




\noindent\textbf{Overlay-based accelerators:}
Overlay accelerators leverage a single generic bitstream to compute across various workloads. They handle different workloads by sending different configuration bits or instructions to control the accelerator performing different operations.
On the one hand, an overlay can use a single bitstream to serve a large set of applications, improving portability and reducing the overhead of reconfiguring the platform.
On the other hand, overlay may lose the opportunity to customize for a special data movement or compute pattern in a certain application, and the capability of serving a wide range of applications could be limited.

A series of overlay accelerators have been proposed for different domains~\cite{RSN,tong2024feather,yang2025nsflow,he2025intar,abdelfattah2018dla,zhang2022fast, BeyondPeakPerformance, Cloud-Scale,guo2024overlay,wei2018tgpa,gemmini-dac}.
AMD Vitis DPU~\cite{amd_vitis_ai_dpu} partitions the on-chip resources to create multiple homogeneous accelerators.
FEATHER~\cite{tong2024feather} leverages a special PE array and a multiple-stage reduction framework, enabling the change of the data layout and data movement for different DNN kernels.
InTAR~\cite{he2025intar} targets the data volume variance in modern DNN workloads, switching execution patterns based on a static schedule to improve computation efficiency and reduce off-chip communication volume.
RSN~\cite{RSN} provides a novel ISA abstraction, modeling the datapath as a circuit-switched network
and enables flexible allocation of heterogeneous functional units.
More recently, FILCO \cite{filco} and DORA \cite{dora} further proposed a more flexible architecture design as well as an ISA abstraction for diverse workloads.
The overlays are not limited to DNN workloads.
For example,~\cite{MonteCarlo} proposes a dataflow overlay for Monte Carlo simulations.

In addition to designing the accelerator architecture, there are also frameworks discussing how to schedule different tasks on overlays.
For example, ~\cite{kim2023moca, zeng2022serving,gao2023layer,oh2021layerweaver, wang2023cd} focus on general scenarios that optimize for the Quality of Service metrics.
~\cite{ji2025art,ji2026derca,restuccia2021time,guan2024mesc} targets different deadline-aware real-time scenarios, models the system in various real-time system setups, and pairs the system design and scheduling with mathematical real-time theory analysis.


\noindent\textbf{Customized accelerators:}
On the other end of the generalization-specialization spectrum is to customize the accelerator architecture for one application and conduct multi-tasking by reconfiguring the FPGA when switching tasks.
Each task can maintain the high efficiency of dedicated hardware design. However, the latency for reconfiguring the FPGA could be a large overhead.
Moreover, it could be costly to enable swapping out an ongoing application and swapping one with higher priority (i.e., preemption), as this requires not only reconfiguring the chip, but also storing back all the on-chip context before configuration.

Multiple existing frameworks customize the accelerator for systems dedicated to one single application.
CHARM\cite{zhuang2023charm}, AutoMM~\cite{zhuang2023automm}, and CHARM 2.0~\cite{charm2} create multiple heterogeneous accelerators to handle the shape variance in the model.
SSR~\cite{ssr} and EQ-ViT~\cite{eq_vit_tcad} target small DNN models and fit the whole model on-chip to reduce the latency and improve throughput.
HMix~\cite{thallikar2026hmix} optimizes quantized DNN inference.
H2PIPE~\cite{doumet2024h2pipe} optimizes the memory bandwidth for devices equipped with HBM.
For non-ML workloads, AIM~\cite{yang2023aim}, PEGASO~\cite{PEGASO}, and Menzel et al.~\cite{Menzel} focus on large number multiplication, stencil kernels, and tensor algebra, respectively.
In addition, several works also focus on the programming abstraction that can improve productivity on heterogeneous accelerator design, including ARIES \cite{aries}, Allo \cite{chen2024allo}, MLIR-AIR \cite{MLIR-AIR},
and IRON \cite{hunhoff2025iron}.

For multi-tasking, there are also works leveraging partial or full reconfiguration to serve different applications.
~\cite{jozwik2010novel,Rossi2018Preemption,attia2020feel} optimize the preemption overheads in reconfiguration.
FRED~\cite{biondi2016framework,pagani2017linux}
proposes a framework for real-time systems using partial reconfiguration.
DART~\cite{seyoum2021automating} leverages the dynamic partial reconfiguration, optimizing for the partitioning and floor planning problem.
UPaRC~\cite{UPaRC}, AC\_ICAP~\cite{cardona2015ac_icap}, MiCAP~\cite{MiCAP}, and VERSATILE~\cite{VERSATILE} optimize for the partial reconfiguration controller and the speed.

\section{Experiment Setup}
\label{sec: setup}

In this section, to evaluate the trade-off between overlay-based execution and customized architectures in realistic autonomous-driving scenarios, we construct an experimental setup inspired by Industry Challenge \cite{industrychallenge}.

\noindent \textbf{Autonomous Driving System (ADS) Setup:}
We adopt a real-time task model to describe the periodic execution of autonomous-driving workloads. An ADS scenario consists of multiple periodic tasks to perceive the environment at an interval. As shown in Figure \ref{fig: sota_timeline}, three tasks are running for environment perception: image segmentation, classification, and point-cloud classification. Depending on the application requirements, each task has a different sensor capture frequency: 20Hz, 15Hz, or 10Hz. Each task instance must finish execution before the next instance of the same task is released. Otherwise, unprocessed sensor data may accumulate, creating a backlog that can eventually violate real-time requirements and compromise system operation. Therefore, the deadline (upward arrows) for each instance is the release time (downward arrows ) of the next instance from the same task. We assume that each accelerator would begin execution as soon as a task instance arrives, and the execution time is the task latency running on the targeted architecture. In addition, following ART \cite{ji2025art}, given the limited computing resources available in embedded systems, multiple tasks can be scheduled on one computing platform by applying the Earliest Deadline First (EDF) scheduling algorithm, prioritizing instances according to their deadlines. 
For example, at 30ms in Figure \ref{fig: sota_timeline}(a), Tasks 1 and 2 are both released and need to be scheduled on the computing platform, since Task 1 has a deadline earlier than Task 2, Task 1 is scheduled. Until Task 1 is completed, Task 2 has the earliest deadline and is scheduled to execute.

\noindent \textbf{Application Setup:}
Based on this real-time task model, we construct workloads from three representative autonomous driving applications, including DeiT for image segmentation, MLP-Mixer for classification, and PointNet for point-cloud classification. Each application is associated with two DNN model variants: one large model and one small model. This gives six models in total, including three large models and three small models. This design is coherent with a common deployment pattern in ADS scenarios, where the system switches between different model sizes depending on scene complexity, accuracy requirements, and available compute resources. In addition, we choose four experiment setups among the combinations of these models, as shown in Table \ref{tab: setup}.

\begin{table}[t]
\centering
\caption{Experimental setup configurations.}
\vspace{-5pt}
\label{tab: setup}
\resizebox{\columnwidth}{!}{%
\begin{tabular}{|c|c|c|c|c|c|c|}
\hline
        & DeiT-S & DeiT-L & MLP-Mixer-S & MLP-Mixer-L & PointNet-S & PointNet-L \\ \hline
Setup A &        & 20Hz   &             & 15Hz        &            & 10Hz       \\ \hline
Setup B & 20Hz   &        &             & 15Hz        &            & 10Hz       \\ \hline
Setup C & 20Hz   &        &             & 15Hz        & 10Hz       &            \\ \hline
Setup D & 20Hz   &        & 15Hz        &             & 10Hz       &            \\ \hline
\end{tabular}%
}
\vspace{-10pt}
\end{table}

\noindent \textbf{Architecture Setup:}
For each experiment setup, we compare two architectural execution strategies. The first strategy uses an overlay architecture \cite{filco}, where all models are handled by an overlay accelerator. Model switching is managed by lightweight instruction and word-level configuration updates, without reloading the bitstream. This strategy provides flexibility and low switching overhead, but may introduce performance degradation compared with fully customized designs. 
The second strategy uses customized architectures, where each model is executed on a model-specific bitstream. This approach can improve execution efficiency for individual models, but switching between different bitstreams may require reloading. Therefore, the total execution cost includes both computing latency and reload overhead whenever consecutive jobs use different customized bitstreams. For customized architectures, we leverage customization architecture, i.e., CHARM \cite{charm2} and SSR \cite{ssr}, as the state-of-the-art hardware design. 
In addition, the reconfiguration latency for customized architecture can be bounded by not only the bandwidth of loading the bitstream from DDR, but also the speed at which the controller configures the system.
For example, when using the AMD Versal VCK190 platform and leveraging the Xilinx Runtime (XRT) to change the bitstream from the embedded CPU processor, a bitstream of size 87MB can take a 25ms reconfiguration latency, whereas the bitstream itself only consumes about 4ms to transfer from DDR to on-chip.
Therefore, we profile the bitstream reload latency on the Versal VCK190 platform with Vitis 2025.2 and Vivado 2025.2, and we set 20 ms as the bitstream reload overhead for the state-of-the-art customized architecture.
\section{Overlay-Customization Tradeoff with state-of-the-art techniques}

\begin{figure}
    \centering
    \includegraphics[width=0.85\linewidth]{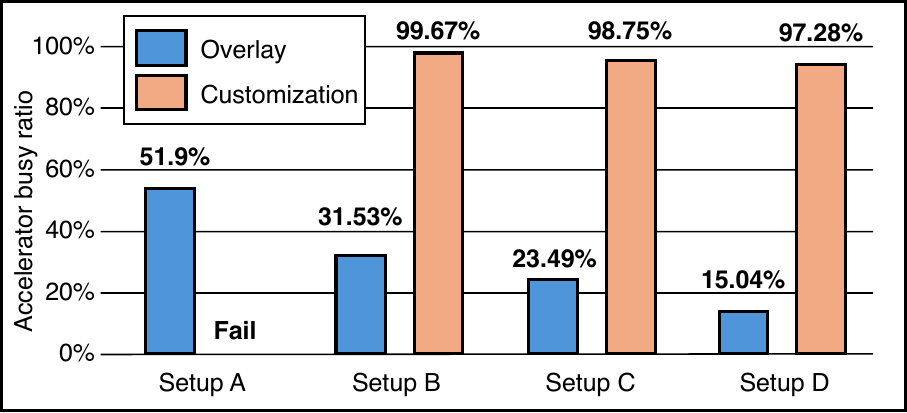}
    \vspace{-10pt}
    \caption{The comparison between overlay and customization on the accelerator busy ratio.}
    \vspace{-10pt}
    \label{fig: sota_util}
\end{figure}

In this section, we discuss the tradeoff between overlay-based and customization accelerators based on state-of-the-art accelerator architectures.
We first inspect the execution timeline on Setup A (Figure \ref{fig: sota_timeline}), then generalize our observation on various setups with heterogeneous applications (Figure \ref{fig: sota_util}).

We evaluate the state-of-the-art accelerators on a task set derived from the real-world perception stack in the Autoware autonomous driving framework \cite{autoware}.
And the execution timeline is depicted in Figure~\ref{fig: sota_timeline}.
When only comparing the execution time, the customized accelerator does present a significant speed up, reaching 20\%, 40\%, and 5\%, compared with overlay accelerator.
However, when serving multiple tasks, every time the accelerator switches tasks, a huge overhead is paid, which is even longer than the execution time.
As a result, the execution time plus reconfiguration overhead is too long to finish each instance before the next one is released.
Unprocessed instances thus accumulate, and for each instance, the response time keeps increasing as time goes on.
As a result, the system fails with this unbounded perception latency.
On the other hand, with the overlay architecture, even though the execution time is higher, without the reconfiguration overhead, the accelerator is able to finish every instance before the next one is released, and can sustain the perception stack.

We then compare the performance in various task sets, as shown in Figure~\ref{fig: sota_util}.
We report the accelerator busy ratio measured in an EDF-based simulation.
For the same task set and same platform, a lower busy ratio suggests that the system has a better computation capability and is able to serve more tasks, or to further scale up the frequency for a more intensive task set. For the setup with task being accumulated (e.g., as that in Figure~\ref{fig: sota_timeline}(a)), we tag them with \texttt{Fail} to suggest they cannot be deployed with unbounded response time.
As shown in Figure~\ref{fig: sota_util}, when overlay is used, the system can run successfully with all four setups.
The highest busy ratio is just above 50\%, suggesting that the system is able to serve more tasks or tasks with smaller overheads.
However, for the customized accelerator, the reconfiguration overhead could be large, so that the unprocessed jobs accumulate in setup A, making the system fail.
For the other three setups, although the system can serve the tasks, the busy ratio is quite high, such that the system cannot serve any more tasks.

It is worth noting that the tradeoff between overlay and customization is also the absolute execution length of the tasks, which is because customization usually brings a latency reduction in the percentile, whereas the reconfiguration overhead is relatively fixed.
As a result, the model must be large enough so that customization can have enough gain to outweigh the overhead.
For example, if a task takes 1 second to finish, even 10\% gain in execution length gets 100 ms, which is much higher than the 20 ms overhead.
Still, for scenarios with low-latency requirements like autonomous driving, as the tasks are usually small, the overlay will consistently be better.

\vspace{-10pt}
\section{Sensitivity Study for Advanced Reconfiguration Techniques}
\label{sec: Advanced Reconfiguration}

Recently, a series of advancements have been proposed towards the reconfiguration controller, or the FPGA architecture, to improve the reconfiguration speed or overlap the reconfiguration latency with execution time \cite{VERSATILE,zhang2025pd}, and some works focus on low-latency context-switching on FPGA, which can reach less than 1 ns \cite{Ferroelectric}.
To demonstrate the potential effects of these techniques, we conduct a sensitivity study to discuss how a smaller reconfiguration overhead may flip the overlay-customization design choice.

\begin{figure}
    \centering
    \includegraphics[width=0.8\linewidth]{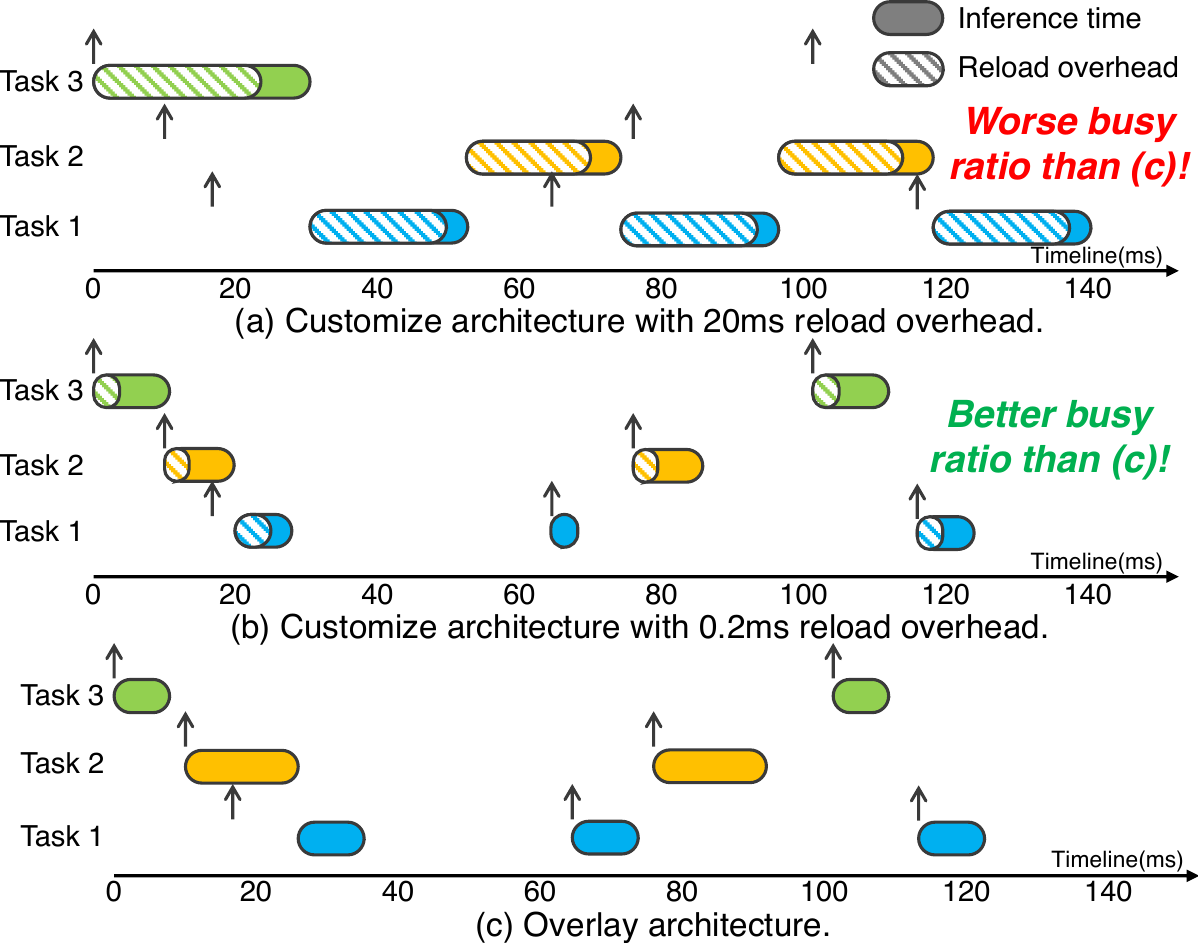}
    \vspace{-10pt}
    \caption{The timeline comparison on customized bitstream with different reload overhead and overlay architecture strategies for Setup B.}
    \vspace{-10pt}
    
    \label{fig: advanced_overhead_timeline}
\end{figure}

\begin{figure}
    \centering
    \includegraphics[width=0.8\linewidth]{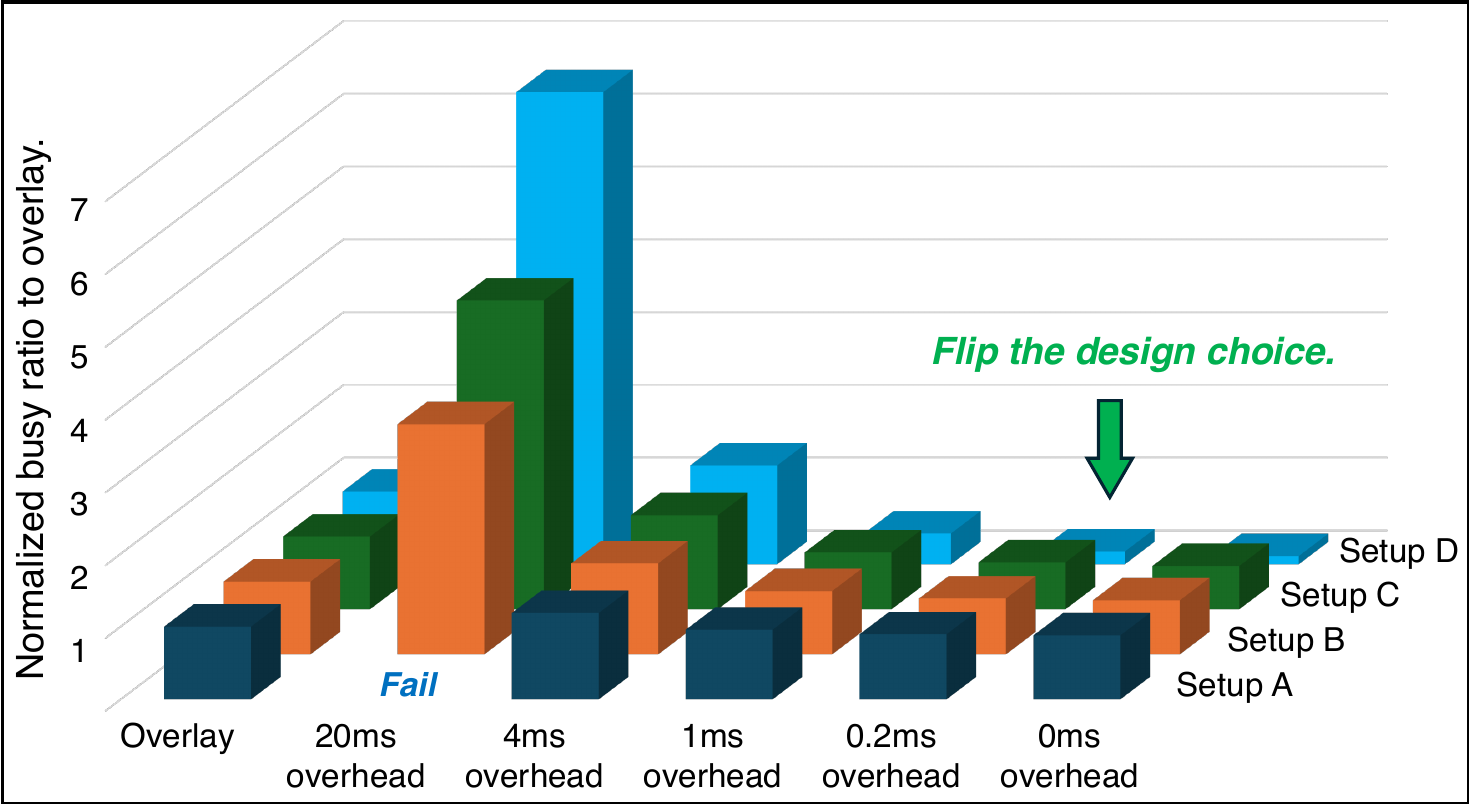}
    \vspace{-10pt}
    \caption{The comparison between overlay and customization with different reload overhead on the accelerator busy ratio. The reload overhead for Cust-1, Cust-2, Cust-3, and Cust-4 is 20ms, 4ms, 1ms, 0.2ms, and 0ms, respectively.}
    \vspace{-15pt}
    \label{fig: bar_advanced_reconfig_exp}
\end{figure}

Figure~\ref{fig: bar_advanced_reconfig_exp} shows the busy ratio of the overlay and customized architecture.
The height is normalized with the overlay's busy ratio.
To represent the improvement of reconfiguration latency, we study 5-levels of different reconfiguration overheads, which are 20ms (state-of-the-art), 4ms, 1ms, 0.2ms, and 0ms, respectively.
The 0ms represents the theoretical upper bound we can achieve with only improving the reconfiguration overhead.

As shown in Figure~\ref{fig: bar_advanced_reconfig_exp}, comparing different levels of overhead, the customized accelerator has a much higher busy ratio when the overhead is large.
As the overhead becomes smaller, the busy ratio also decreases, and finally, when the overhead is 1 ms, the choice is flipped:
For all four setups, the customized accelerator will have a smaller busy ratio compared with the overlay, reaching 0.96$\times$, 0.85$\times$, 0.79$\times$, and 0.42$\times$ of the overlay busy ratio.
Compared with 20ms state-of-the-art overhead, this suggests that a roughly 20$\times$ reduction is needed in the reconfiguration overhead to enable the customized accelerator to outperform overlay.

When comparing different setups, we can find that the task set with more small models is more sensitive to the change of reconfiguration overhead.
For Setup B with two large and one small models, switching the overhead from 4 to 1 ms changes the busy ratio from 1.19$\times$ to 0.96$\times$ of the overlay architecture.
On the other hand, for Setup D, where all tasks are small ones, this reduction is from 1.35$\times$ to 0.42$\times$.
The reason is twofold: First, as the execution time for these models is smaller, the fixed reconfiguration overhead will take a larger proportion, incurring a higher normalized busy ratio.
On the other hand, the small models can have a larger improvement through customization techniques such as on-chip forwarding. When the overhead is small, the small models can have a larger reduction in execution time and get a lower busy ratio.

This effect can be observed straightforwardly in Figure~\ref{fig: advanced_overhead_timeline}.
As shown in Figure~\ref{fig: advanced_overhead_timeline}(b), after reducing the reconfiguration overhead, the system can almost run and finish the job immediately after it is released.
Comparing Figure~\ref{fig: advanced_overhead_timeline}(b) and Figure~\ref{fig: advanced_overhead_timeline}(c), with the optimized execution time and reduced overhead, the total execution length in the customized accelerator now outperforms overlay in task 1 and task 2, and is slightly higher in task 3.
As a result, the overall busy ratio for the customized accelerator is lower than the overlay.

\vspace{-10pt}
\section{Case Study for Advanced Overlay Architecture Design}

Similar to the advancements in reconfiguration overheads, the overlay architecture is also evolving.
In this section, we also analyze the impact of potential advancements in overlay on the ADS scenarios using a case study.
As discussed in Section~\ref{sec: related works}, one limitation of overlay is that the execution pattern is relatively fixed for tasks with various shapes.
Thus, for small tasks whose parameter and activation size are small enough to fit on-chip, its execution latency on an overlay could be suboptimal when a kernel-wise execution pattern is used, which loads and stores from and to DDR in every kernel, incurring a large amount of memory latency.
Through architectural advancement and novel instruction set architecture (ISA) abstraction, it is possible to improve the memory movement of the overlay architecture, allowing moving the data between different on-chip buffers so that the inter-layer intermediate data can be forwarded on-chip.

\begin{figure}
    \centering
    \includegraphics[width=0.8\linewidth]{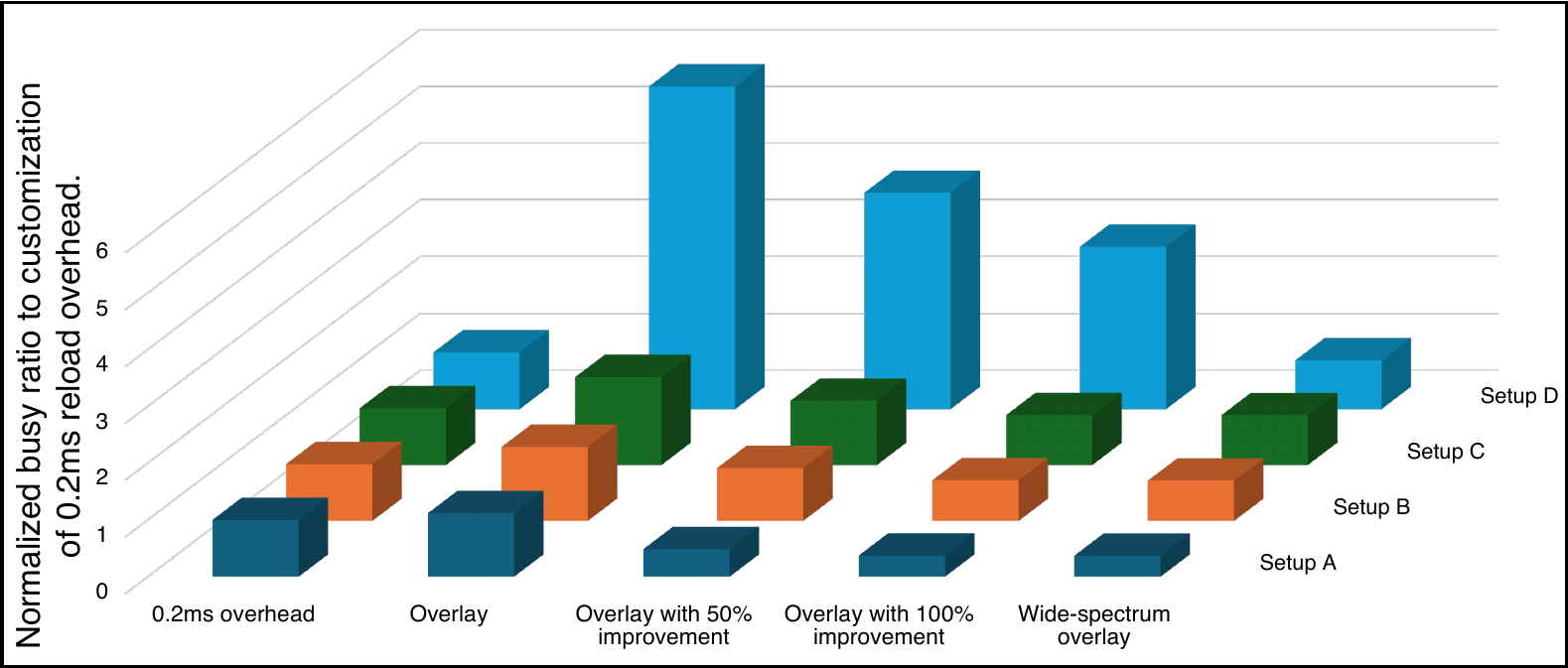}
    \caption{The comparison between customization and overlay with different potential throughput improvement on the accelerator busy ratio. The throughput improvement of the overlay architecture is 50\%, 100\%, and the wide-spectrum overlay architecture, respectively.}
    \vspace{-15pt}
    \label{fig: advanced_overlay_exp}
\end{figure}

Figure \ref{fig: advanced_overlay_exp} shows the sensitivity study, and the baseline customized architecture with a 0.2ms reload latency is normalized to 1.
The difference is that for the overlay architecture, we assume that the original architecture can be further improved by enabling on-chip forwarding and proposing a more advanced ISA design, even a wide-spectrum overlay that can handle diverse shapes, and also cover models that fit on-chip, and not fit on-chip efficiently. Under the current overlay design, the busy ratio is higher than the customized baseline across all setups, especially for the lightweight workload in Setup D, where the overlay busy ratio reaches 5.68$\times$. 
This indicates that the current overlay still suffers from non-negligible low efficiency when handling small models.
However, as the throughput of the overlay is improved, the design choice changes significantly. With a 50\% throughput improvement, the overlay becomes competitive or better in Setup A and Setup B, while the gap is substantially reduced in Setup C and Setup D. With a 100\% throughput improvement, the overlay outperforms the customized baseline in Setups A, B, and C, reducing the busy ratio to 0.37$\times$, 0.72$\times$, and 0.88$\times$, for Setup A to C, respectively. 
For Setup D, since all three models are small, customized architectures such as SSR \cite{ssr} can achieve substantial throughput gains from on-chip forwarding and kernel fusion optimization.
Prior customized accelerators, such as SSR \cite{ssr} and CHARM \cite{zhuang2023charm}, require dedicated bitstreams for different models, while state-of-the-art overlay architectures, FILCO and DORA, take a step toward a wider overlay by enabling flexible dataflow. However, the case study indicates that it requires a wide-spectrum overlay covering both CHARM-style fit on-chip models and SSR-style not fit on-chip models to maintain both high flexibility and low overhead.

\vspace{-10pt}
\section{Conclusion}

This paper studies the deployment tradeoff between overlay-based and customization-based architectural design for ADS scenarios. 
The results show that under state-of-the-art reconfiguration latency, overlays are more suitable for frequent model switching since they can avoid costly bitstream reload and maintain high flexibility. 
However, this tradeoff depends on the technique improvement: faster reconfiguration can favor customized architectures, while more capable overlays can shift the advantage back. 

\smallskip
{\small
{\noindent\textbf{ACKNOWLEDGEMENTS --}} This work is supported in part by Brown University New Faculty Start-up Grant, DOE award DE-SC0026344,
NSF awards 
\#2140346, 
\#2231523, 
\#2441179, 
\#2348306, 
\#2511445, 
\#2518375, 
\#2536952, 
\#2544032. 
We thank AMD for the hardware and software donations.
P. Zhou has a financial interest in Shanghai Suikun.
\vspace{-10pt}

\bibliographystyle{ACM-Reference-Format}
\bibliography{Ref_X}

\end{document}